%% file: Draft.tex
\documentclass[aps,reprint,twocolumn,superscriptaddress]{revtex4-2}

\usepackage[caption=false]{subfig}
\usepackage[justification=justified]{caption}
\usepackage{float}
\usepackage{mathtools}
\usepackage{bbold}
\usepackage{amsmath,amsfonts,amsthm,braket}
\usepackage{adjustbox}
\usepackage{physics}
\usepackage{graphicx}% Include figure files
\usepackage{dcolumn}% Align table columns on decimal point
\usepackage{bm}% bold math
\usepackage[normalem]{ulem}

\begin{document}

%\preprint{APS/123-QED}

\title{Linking quantum discord with Bayesian game theory}% Force line breaks with \\

\author{Adam Lowe}
 \email{adam.j.lowe90@gmail.com}
\affiliation{%
Department of Applied Mathematics and Data Science, Aston University
}%

\date{\today}% It is always \today, today,

\begin{abstract}
Witnessing quantum correlation in real, practical experimental set-ups is a key focus for the development of near-future quantum technologies. Whilst there are several experimental protocols for witnessing entanglement, detecting quantum discord remains far more elusive. A recently proposed quantum discord witness offers an experimentally accessible set-up which allows the quantum discord to be witnessed through a non-linear combination of correlation functions. Interestingly, the experimental set-up can be mapped to Bayesian game theory allowing for an extended generalisation of the proposed witness. Subsequently, it is shown that there is a direct link between the expected payoff in Bayesian game theory and the previously proposed quantum discord witness by uniting these two concepts through the established CHSH game.
\end{abstract}

\maketitle

%\tableofcontents

\section{Introduction}
Since the inception of quantum mechanics \cite{bell}, there has been extensive work on how to witness and quantify quantum effects \cite{entanglement_toth,morris_yadin} due to their promising technological capabilities. In particular, determining whether these quantum effects could be utilised to demonstrate an advantage of using quantum physics over classical physics \cite{quantumadvantage} has been a driving motivation, due to their promise to enhance modern-day technological capabilities. The main resource for quantum advantage is due to quantum entanglement and non-locality as these offer clear advantage over using classical physics. Whilst there has been substantial theoretical and experimental development in understanding fundamental quantum correlations, primarily focusing on entanglement, there is still debate on how best to witness quantum correlations beyond entanglement, namely quantum discord \cite{discord,discord1}. This is since quantum discord can capture local quantum correlation, whereas most classifiers of quantum correlation rely on differentiating between local and non-local states. An added benefit of quantum discord is due to its relative robustness against external perturbations and noise \cite{Werlang:2009aa,:2012aa} compared to other forms of quantum correlation, ensuring it will have practical relevance for the development of future quantum technologies.

The main experiment which highlights the difference between quantum and classical physics utilises the Bell inequalities and extends them into an experimentally accessible form through the CHSH inequality \cite{PhysRevLett.23.880}, which is built from a linear combination of correlation functions in a bipartite system. Crucially, violation of the CHSH inequality can only be achieved when the two parties share a non-local quantum state \cite{aspect}. This is the most well-known experiment which can demonstrate how quantum correlation can be advantageous relative to classical correlation (sometimes termed as shared randomness).

Whilst the CHSH inequality works well as a witness for non-local quantum correlation, it fails as a witness for quantum discord, as discorded states can be local. Therefore, the CHSH inequality only distinguishes between locality and non-locality allowing some quantum (mixed-separable) states not to be identified. This is an issue when trying to design and develop quantum technologies, as understanding the type of quantum correlation allows suitable algorithms and protocols to be designed in order to utilise the quantum advantage. Therefore this motivates a clear need to design an experimentally friendly witness of quantum discord which allows for clear quantification of quantum effects \cite{High_Energy_QD}, which exist beyond non-locality.

In recent years, techniques from other fields of science, in particular information theory have been used as a promising candidate to allow for the detection and quantification of quantum correlations. This has enabled the nascent field of quantum information theory to develop, both theoretically and experimentally demonstrating tremendous success \cite{igor_paper,lowe,disc_2qbit,IP,2byN_discord}. Given the success in utilising other aspects of mathematics, it is not surprising that over the past 25 years, quantum correlations have been incorporated into game theory \cite{nasheq,QgameRev,PhysRevA.65.022306,1999PhRvL,Ikeda:2023aa,PhysRevLett.82.1052,Ikeda:2023ab,Iqbal2000EvolutionarilySS,Lowe:2024aa,Cheon:2008aa,Guha2021quantumadvantage,Andronikos:2022aa} allowing a clear operational framework in which to devise practical experiments. Since game theory studies conflict and cooperation between interacting parties on a network \cite{vonneumann1947}, it has clear relevance for the development of future quantum technologies as the optimal design of quantum networks is necessary for the successful implementation of quantum advantage in modern society through novel protocol design.  Moreover, since the fundamental mathematical framework for game theory has already been developed, there are clear established techniques which can allow for quantum advantage to be witnessed.

The advent of quantum game theory initially incorporated entangled states and demonstrated how non-local unitary operations could be utilised as a strategic mechanism demonstrating clear quantum advantage relative to the classical counterpart which would be limited by local unitary operation \cite{Qgame_Locality}. There was initial success by demonstration of quantum advantage compared to the known classical optimal solutions in previously studied games. However, there were questions about how to {\it{play}} a non-local unitary strategy. Subsequently, to avoid any philosophical questions about how to play these types of games, different game theory techniques have been developed to replicate the standard experimental set-up for the CHSH inequality which makes use of two separated parties performing repeated measurements on a shared state \cite{bellbayes}. It is this framework which forms the basis of the analysis conducted in this paper. This is formally introduced below.

\section{Bayesian Game Theory}
\label{BayesG}
The advent of Bayesian game theory \cite{harsanyi} yields a general mechanism to model how parties interact and formulate decisions when there is uncertainty in what type of situation they are in. This is a realistic scenario which arises in everyday events from auctions to traffic management \cite{Farooqui:2016aa}. This notion of the type of game is crucial, as it is the type which is where the uncertainty lies. For example, in an auction scenario, there may be two bidders, one who is risk-averse, and one who is more likely to take risks. Crucially both bidders are unaware about the type of the other bidder, resulting in each player having uncertainty about which type of player they are playing against. This results in the Bayesian nature of these games as each player has to assign a prior belief about what type of player the other player is, based on what they believe is the likely outcome. This is a powerful formalism as it mirrors many real-world scenarios where uncertainty, or lack of knowledge, is a fundamental limitation. 

Recently, Bayesian game theory has been utilised as a mechanism which allows a general framework in which to formulate Bell-type experiments where there are two parties each performing measurements in their own respective subsystems \cite{Cleve:2004aa}. Based on these measurements, they can assign numbers termed {\it{payoffs}} for each result which obey some preference relation. For example, suppose that Alice and Bob both measure spin up, they can assign the payoff ``1''. It is important to stress, that these payoffs are completely arbitrary, so there is complete freedom over what they are chosen to be, however the player's decisions are dictated by the preference relations as each player wishes to choose their strategies in order to be in an optimal situation. This is where the concept of the Nash equilibrium originates from. For simplicity, in the rest of this paper it is assumed that the largest payoff is preferred by the players, namely $1$ is preferred over $0$. To clarify the experimental procedure, two parties (termed Alice and Bob) are each sent a copy of a shared state which they then perform judicious measurements on. Once they measure the outcome of their measurement (for example, a particle either being spin up or spin down), they send the measurement outcome to an independent referee to collate the results, and assign a given payoff based on the outcome. Crucially, each player can be in two different types of game, however, they don't know beforehand which type of game they will be in, nor which type of the game the other player will be in. This allows a Bayesian formulation of the setup. This experiment is repeated many times to allow an average payoff to be determined. The objective of a cooperative Bayesian game is to maximise the average (expected) payoff.

The general definition of the expected payoff for player A for the incomplete information (Bayesian) game described above is given by
\begin{equation}
\label{expPay}
u_A = \sum_{\sigma,\sigma', {\bm{\alpha}},{\bm{\beta}}} P_A({\bm{\alpha}},{\bm{\beta}}) P(\sigma \sigma' | {\bm{\alpha \beta}}) u_{\sigma,\sigma',A}^{{\bm{\alpha}},{\bm{\beta}}},
\end{equation}
where $P_A({\bm{\alpha}},{\bm{\beta}})$ is player A's joint prior belief, $u_{\sigma,\sigma',A}^{{\bm{\alpha}},{\bm{\beta}}}$ is player A's tensor of payoff's, and 
\begin{equation}
P(\sigma \sigma' | {\bm{\alpha}} {\bm{\beta}}) = \tr \Big[ \Pi_{\sigma | {\bm{\alpha}}} \otimes \Pi_{\sigma' |  {\bm{\beta}}} \rho \Big],
\end{equation}
with
\begin{equation}
\Pi_{\sigma | {\bm{\alpha}} }= \frac{1}{2}(\mathbb{1}+\sigma {\bm{\alpha}} .{\boldsymbol{\sigma}}),
\end{equation}
where $\rho$ is the state of the system, $\boldsymbol{\sigma}$ is the vector of Pauli matrices, $\sigma,\sigma' = \{\uparrow,\downarrow\} = \{+,-\}$, ${\bm{\alpha}} = \{{\bf{a}},{\bf{a'}} \}$, and ${\bm{\beta}} = \{{\bf{b}},{\bf{b'}} \}$. Note $\sigma,\sigma'$ are spin measurements are relate to the player's measurement outcomes, and ${{\bm{\alpha}},{\bm{\beta}}}$ are the types of games that the players can be in. Additionally ${\bf{a}}$ is parameterised on the Bloch sphere such that  ${\bf{a}} = (\sin \theta_a, 0, \cos \theta_a)$, similarly for ${\bf{a'}},{\bf{b}},{\bf{b'}}$. Note, the spherical angle, $\phi_a$ has been set to zero for simplicity. 
To emphasise the interpretation of Table I, the players are unaware of which type of game they will be, for example player A is unaware whether they are in type $\bf{a}$ or $\bf{a'}$. Similarly they are unsure if player B is in type $\bf{b}$ or $\bf{b'}$. The same reasoning applies for player B. Based on what each player measures (either spin up or spin down), and depending on what type of game they are in, they will be assigned a payoff given in Table I. For example, suppose player A is in type $\bf{a}$ and player B is in type $\bf{b}$, and they both measure spin up, then they will each be assigned a payoff of $1$. It should be clarified that the player's do not necessarily have to be assigned the same payoffs, for example in a zero-sum game, one player wins at the expense of the other. However, adversarial games are beyond the scope of this paper.

It is well established, that by using non-local states, where non-locality is defined as
\begin{equation}
\rho^{AB} \neq \sum_i p_i \rho_{i}^{A} \otimes \rho_{i}^{B},
\end{equation}
the maximum classical expected payoff can be surpassed. This is related to the conditional probability not being able to be split into a product of joint probability distributions, i.e. it is not separable and thus can not be described using local hidden variables. The question remains, whether this framework allows the possibility of witnessing quantum correlations which are not entangled. It is clear that since quantum discord can exist for mixed separable states, violation of the maximum classical bound could only be strictly due to non-locality. However, it is possible that there may be some generic relation between the expected payoff (which takes into account, the priors, the shared state between the players, and the tensor of actual payoffs) and the quantum discord. To further motivate this question, quantum discord is introduced formally below.

\section{Quantum Discord and Witness}
Quantum discord allows quantum correlations to be revealed which exist when the state may be unentangled. It is an information-theoretic measure which qualitatively is the minimised difference between the mutual information and the quantum mutual information and has been proposed as a useful resource for cryptographic purposes \cite{Pirandola:2014aa}. The mutual information is given by 
\begin{equation}
I(\rho^{AB}) = S (\rho^A) + S(\rho^B) - S(\rho^{AB}),
\end{equation}
where $S(\rho^{AB}) = - \tr \rho^{AB} \log \rho^{AB} = - \sum_i \lambda_i \ln \lambda_i$, $\rho^{AB}$ is the density matrix of a bipartite system, and $\lambda_i$ are the eigenvalues of the density matrix. 
The quantum mutual information is given by
\begin{equation}
J_A (\rho^{AB}) =   S(\rho^A) -  S(A | \Pi_{\sigma}^{B}),
\end{equation}
where $S(\rho^{A}) = - \tr_B \rho^{AB} \log \rho^{AB}$, $\sigma=\pm 1$ analogous to $\{\uparrow,\downarrow\}$, $S(A | \Pi_{\sigma}^{B}) = \sum_{\sigma} p_{\sigma} S(\rho_{A | \Pi_{\sigma}^{B}} )$, and
\begin{equation}
\rho_{A | \Pi_{\sigma}^{B}} = \frac{1}{p_{\sigma}} \tr_B (\mathbb{1} \otimes \Pi_{\sigma}^{B}) \rho^{AB}  (\mathbb{1} \otimes \Pi_{\sigma}^{B}),
\end{equation}
with $p_{\sigma} = \tr  (\mathbb{1} \otimes \Pi_{\sigma}^{B}) \rho^{AB}$, $\Pi_{\sigma}^{B} = \frac{1}{2}(\mathbb{1}+\sigma \bf{n}.\boldsymbol{\sigma})$, where $\bf{n}$ is the Bloch vector corresponding to $\{{\bf{a}},{\bf{a'}},{\bf{b}},{\bf{b'}}\}$ and $\boldsymbol{\sigma}$ is defined as before.
This allows the quantum discord to be given by
\begin{equation}
\begin{split}
D_A (\rho^{AB}) &=  \min\limits_{\Pi_{\sigma}^{B}} [ I(\rho^{AB}) - J_A (\rho^{AB})] \\&= \min\limits_{\Pi_{\sigma}^{B}} S(A | \Pi_{\sigma}^{B}) + S(\rho^B) - S(\rho^{AB}).
\end{split}
\end{equation}
It should be noted that quantum discord is an asymmetric relation \cite{Jebarathinam:2023aa}, as the measurement can instead be performed on subsystem $A$ and optimised over. Whilst this gives a general relation of quantum correlation, its practical use is limited as it requires full knowledge of the density matrix which in general requires full quantum state tomography, in addition to performing minimisation over all projective measurements. Therefore finding an experimentally accessible witness of quantum discord is of paramount importance. Fortunately, a recent result \cite{disc_WITNESS} devised a non-linear witness in terms of correlation functions given by
\begin{equation}
\label{witness}
W \coloneqq \begin{vmatrix} Q_{{\bf{a}}{\bf{b}}} & Q_{{\bf{a}}{\bf{b'}}} \\ Q_{{\bf{a'}}{\bf{b}}} & Q_{{\bf{a'}}{\bf{b'}}} \end{vmatrix} = Q_{{\bf{a}}{\bf{b}}}Q_{{\bf{a'}}{\bf{b'}}} - Q_{{\bf{a}}{\bf{b'}}}Q_{{\bf{a'}}{\bf{b}}},
\end{equation}
where $Q_{{\bm{\alpha}} {\bm{\beta}}} = \langle A_{\bm{\alpha}} \otimes B_{\bm{\beta}} \rangle - \langle A_{\bm{\alpha}} \rangle \langle B_{\bm{\beta}} \rangle$. Furthermore, the expectation of the players' measurements is defined as
\begin{equation}
\label{expPlayers}
\begin{split}
& \langle A_{\bm{\alpha}} \otimes B_{\bm{\beta}} \rangle = \tr \big( A_{\bm{\alpha}} \otimes B_{\bm{\beta}} \rho^{AB} \big) \\ &= P(\uparrow \uparrow |{\bm{\alpha}} {\bm{\beta}}) + P(\downarrow \downarrow |{\bm{\alpha}} {\bm{\beta}}) - P(\downarrow \uparrow|{\bm{\alpha}} {\bm{\beta}}) - P(\uparrow \downarrow|{\bm{\alpha}} {\bm{\beta}}),
 \end{split}
 \end{equation}
 with player A's expectation value given by $\langle A_{\bm{\alpha}} \rangle = \tr \big( A_{\bm{\alpha}} \rho^A \big) = P(\uparrow |{\bm{\alpha}}) - P(\downarrow|{\bm{\alpha}})$ where player A measures spin up and spin down in game type $\bm{\alpha}$, and player B's expectation value is $\langle B_{\bm{\beta}} \rangle = \tr \big( B_{\bm{\beta}} \rho^B \big) = P(\uparrow|{\bm{\beta}}) - P(\downarrow|{\bm{\beta}})$ for similar measurements and game type $\bm{\beta}$. To ensure the notation in the experimental set up for this witness is the same as described in section \ref{BayesG}, the notation from \cite{disc_WITNESS} has been changed such that $\{x,y\} \rightarrow \{{\bm{\alpha}},{\bm{\beta}}\}$ and $\{0,1\} \rightarrow \{\uparrow,\downarrow\}$. Given the experimental procedure to witness quantum discord can be mapped over to a Bayesian game implies there is a direct link between the two frameworks. This is shown rigorously below.

\section{Linking Quantum Discord Witness to Bayesian Game Theory}
\subsection{Proof}
Table I denotes the available payoffs to each of the players, based on their measurement outcomes and the types of games that they are in.
\begin{table}[htp!]
     \begin{adjustbox}{width=0.3\textwidth}
    \begin{tabular}{c|c c}
     (${\bm{\alpha}},{\bm{\beta}}$)   & ${\bf{b}}$ & ${\bf{b'}}$ \\ \hline
     ${\bf{a}}$  & \begin{tabular}{c|c c}
       &  {$\uparrow$} &{$\downarrow$}\\ \hline
       {$\uparrow$}  & 1 & -1  \\
      {$\downarrow$} & -1& 1
     \end{tabular}  & \begin{tabular}{c|c c}
       &  {$\uparrow$} &{$\downarrow$}\\ \hline
       {$\uparrow$}  & 1 & -1  \\
      {$\downarrow$} & -1& 1
     \end{tabular} \\ 
     ${\bf{a'}}$  & \begin{tabular}{c|c c}
       &  {$\uparrow$} &{$\downarrow$}\\ \hline
       {$\uparrow$}  & 1 & -1  \\
      {$\downarrow$} & -1& 1
      \end{tabular} & \begin{tabular}{c|c c}
       &  {$\uparrow$} &{$\downarrow$}\\ \hline
       {$\uparrow$}  & -1 & 1  \\
      {$\downarrow$} & 1& -1
      \end{tabular}
    \end{tabular}
    \end{adjustbox}
    \caption{This table shows the payoff matrix for an incomplete information game. Based on what bits they receive, from the set of $\bm{\alpha}$ and $\bm{\beta}$, denotes how Alice's and Bob's detectors are set up. From this, they then perform measurements on their shared states and based on their results, they are assigned a payoff. Using this, it is found when Alice's and Bob's measurements are the same, i.e. they both measure $\uparrow$ or $\downarrow$, they agree beforehand to assign a payoff of 1, when in a particular type of game. When Alice's and Bob's measurements differ, for example Alice measures $\downarrow$, and Bob measures $\uparrow$, then they assign a payoff of -1. It is crucial to emphasise that the payoffs differ depending on what type of game the players are in. Again, it is also important to highlight these payoffs in the table can be arbitrary, however they are currently chosen to map directly to the CHSH inequality.}
\end{table}

By now using Eq. (\ref{expPlayers}), and remembering the previously discovered relation between Bayesian game theory and Bell non-locality, it is pertinent to remember that the CHSH inequality is given by
\begin{equation}
\label{CHSH}
\begin{split}
\langle C \rangle &= \langle A_{\bf{a}} \otimes B_{\bf{b}} \rangle + \langle A_{\bf{a}} \otimes B_{\bf{b'}} \rangle + \langle A_{\bf{a'}} \otimes B_{\bf{b}} \rangle \\& - \langle A_{\bf{a'}} \otimes B_{\bf{b'}} \rangle,
\end{split}
\end{equation}
where it has been written in terms of the notation introduced in this paper. By rewriting {\eqref{witness}} in terms of the components of ${\bm{\alpha}},{\bm{\beta}}$ with $W= Q_{{\bf{a}}{\bf{b}}}Q_{{\bf{a'}}{\bf{b'}}} - Q_{{\bf{a}}{\bf{b'}}}Q_{{\bf{a'}}{\bf{b}}}$, and remembering that $Q_{{\bf{a}}{\bf{b}}} = \langle A_{{\bf{a}}} \otimes B_{{\bf{b}}} \rangle  - \langle A_{{\bf{a}}} \rangle  \langle B_{{\bf{b}}} \rangle$ allows \eqref{CHSH} to be written as
\begin{equation}
\begin{split}
\label{CHSH1}
\langle C \rangle &= \frac{W + Q_{\bf{a}\bf{b'}}Q_{\bf{a'}\bf{b}}}{Q_{{\bf{a'}}\bf{b'}}} + \langle A_{\bf{a}} \rangle \langle B_{\bf{b}} \rangle + \langle A_{\bf{a}} \otimes B_{\bf{b'}} \rangle  \\& + \langle A_{\bf{a'}} \otimes B_{\bf{b}} \rangle- \langle A_{\bf{a'}} \otimes B_{\bf{b'}} \rangle.
\end{split}
\end{equation}
The payoffs given in Table I can be mapped directly to $\langle A_{{\bm{\alpha}}} \otimes B_{{\bm{\beta}}} \rangle$ for all ${\bm{\alpha}},{\bm{\beta}}$. This is seen explicitly when both players attain either the outcome $\uparrow$ or both attain the outcome $\downarrow$ in type ${\bf{ab}}$ this gives a payoff of 1, whereas when their measurement outcomes are opposite, they get a payoff of $-1$. This is exactly described by
\begin{equation}
\begin{split}
\langle A_{{\bf{a}}} \otimes B_{{\bf{b}}} \rangle &=P(\uparrow \uparrow| {\bf{a}}{\bf{b}}) +P(\downarrow \downarrow |{\bf{a}}{\bf{b}}) -P(\uparrow \downarrow | {\bf{a}}{\bf{b}}) \\ &-P(\downarrow \uparrow|{\bf{a}}{\bf{b}}) = \sum_{\sigma,\sigma'} P(\sigma \sigma' |  {\bf{a}}{\bf{b}}) u_{\sigma,\sigma'}^{ {\bf{a}}{\bf{b}}},
\end{split}
\end{equation} 
where the subscript $A$ in the tensor of payoff's has been removed as it is cooperative game so the payoff's for $A$ and $B$ are the same, and remembering the mapping between the spins and measurement outcomes. Since there is the subtlety that Table I is a cooperative game, it is also assumed that each player has the same prior belief's about the other player. Clearly, this can be extended for the different types of games. The final consideration is related to the value of the prior belief's. In the CHSH game and the discord witness experiment, the types are chosen randomly by both players, yielding a joint probability distribution given by $P({\bf{a}},{\bf{b}}) =P({\bf{a}},{\bf{b'}}) =P({\bf{a'}},{\bf{b}}) =P({\bf{a'}},{\bf{b'}}) =1/4$.  Therefore, using this information with Table I, and combining \eqref{expPay} and \eqref{CHSH}, it is seen that there is a clear relation between the expected payoff and $\langle C \rangle$ given by
\begin{equation}
\label{link}
u = \frac{1}{4}  \langle C \rangle,
\end{equation}
where $u_A = u_B = u$.
Whilst, the link between Bayesian game theory and the CHSH inequality has previously been reported by Brunner and Linden \cite{bellbayes}, its relevance for finding a direct link between the expected payoff and quantum discord remained undiscovered. By now combining and rearranging \eqref{CHSH1} and \eqref{link}, it is clear there is a direct link between Bayesian game theory and witnessing quantum discord as
\begin{equation}
\begin{split}
u &= \frac{1}{4} \Bigg[\frac{W + Q_{\bf{a}\bf{b'}}Q_{\bf{a'}\bf{b}}}{Q_{{\bf{a'}}\bf{b'}}} + \langle A_{\bf{a}} \rangle \langle B_{\bf{b}} \rangle + \langle A_{\bf{a}} \otimes B_{\bf{b'}} \rangle  \\& + \langle A_{\bf{a'}} \otimes B_{\bf{b}} \rangle- \langle A_{\bf{a'}} \otimes B_{\bf{b'}} \rangle \Bigg].
\end{split}
\end{equation}
This can be further simplified by writing 
\begin{equation}
\label{finalEq}
\begin{split}
u = \frac{1}{4} \Big[ W ( \mu - \nu ) + \eta + \mu Q_{\bf{a}\bf{b'}} Q_{\bf{a'}\bf{b}} + \nu Q_{\bf{a}\bf{b}} Q_{\bf{a'}\bf{b'}} \Big],
\end{split}
\end{equation}
where 
\begin{equation}
\mu = \frac{1}{Q_{\bf{a'}\bf{b'}}} - \frac{1}{Q_{\bf{a}\bf{b}}}, \hspace{0.5cm} \nu = \frac{1}{Q_{\bf{a}\bf{b'}}} + \frac{1}{Q_{\bf{a'}\bf{b}}},
\end{equation}
and 
\begin{equation}
\begin{split}
\eta &= \langle A_{{\bf{a}}} \rangle  \langle B_{{\bf{b}}} \rangle +  \langle A_{{\bf{a'}}} \rangle  \langle B_{{\bf{b}}} \rangle +  \langle A_{{\bf{a}}} \rangle  \langle B_{{\bf{b'}}} \rangle \\&-  \langle A_{{\bf{a'}}} \rangle  \langle B_{{\bf{b'}}} \rangle.
\end{split}
\end{equation}
At this stage, there are a couple of relevant aspects that need to be addressed. It is clear that divergence can emerge if $Q_{{\bf{a'}}\bf{b'}} = 0$. However, remembering that $W= Q_{{\bf{a}}\bf{b}}Q_{{\bf{a'}}\bf{b'}} - Q_{\bf{a}\bf{b'}}Q_{\bf{a'}\bf{b}}$ and noting this was solved for $Q_{{\bf{a}}\bf{b}}$ implies that when $Q_{{\bf{a'}}\bf{b'}} = 0$, it is not possible to solve for $Q_{{\bf{a}}\bf{b}}$ in the first place so the divergence is not realised in practice. This logic applies for other combinations of $Q_{{\bm{\alpha}}{\bm{\beta}}}$. Following on from this, it is seen that there are many different ways to relate the discord witness to the expected payoff, depending on which $Q_{{\bm{\alpha}}{\bm{\beta}}}$ is solved for. However, the essence of the relation will remain. To clarify, it is always possible to re-write Eq. (\ref{finalEq}) such that it takes the form of Eqs. (\ref{CHSH}) and (\ref{link}). This is because $u$ is a functional, where $W$ explicitly depends on $Q_{{\bm{\alpha}}{\bm{\beta}}}$. Therefore, $W$ can always be written in terms of combinations of $Q_{{\bm{\alpha}}{\bm{\beta}}}$. This also explains the exhibited behaviour of a distribution in Figs. 2b and 2c. Ultimately, Eq. (\ref{finalEq}) clarifies how general quantum correlation can arise when performing CHSH-like experiments, rather than focusing solely on entanglement, even if it can not always be detected through violation of inequalities.

Whilst, this result is found for specific payoffs to relate to the CHSH inequality, an entirely generic relation between the expected payoff and the discord witness should be possible. It is also prudent to note that this discord witness is sufficient, but not necessary. Therefore, there are certain discorded states which do not exhibit a non-zero witness. A final aspect is that the discord witness can be both positive and negative. This naturally introduces the question of whether the discord witness increases monotonically with the expected payoff.

\subsection{Examples}
\begin{figure*}[htp]
 % \centering
  \subfloat[a][The expected payoff and discord witness for the Werner state.] {\includegraphics[width=0.47\linewidth]{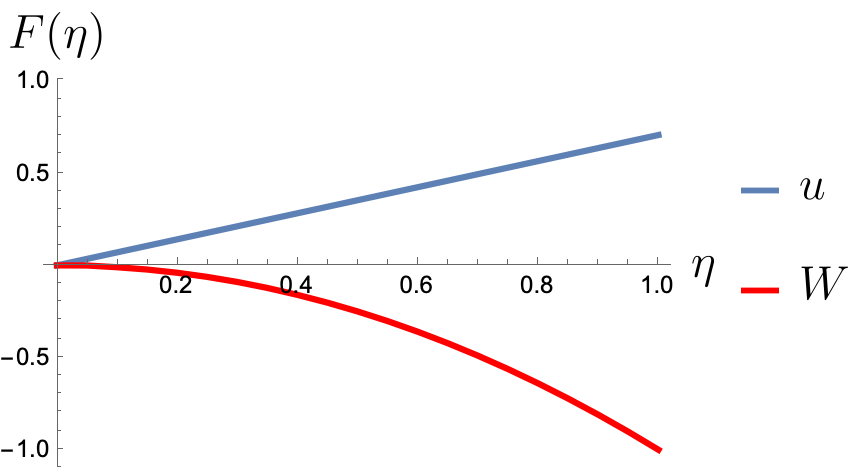}} 
  \label{fig:a} \subfloat[a][The expected payoff and discord witness for $\rho_D(x)$.] {\includegraphics[width=0.47\linewidth]{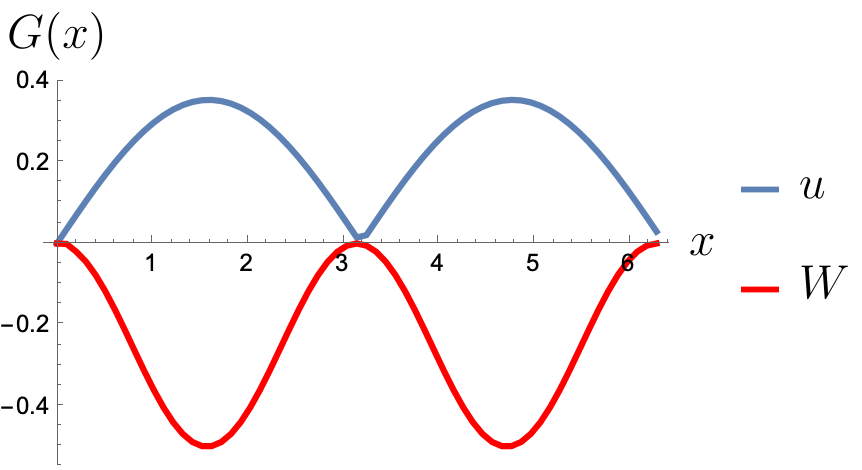}} 
  \label{fig:g}   \caption{This figure shows the expected payoff and discord witness when the expected payoff is maximised. The measurement angles ($\theta_a, \theta_{a'}, \theta_b, \theta_{b}$) which maximise the expected payoff are then substituted into the discord witness to determine how these functions interact. It is clear that the discord witness is non-zero when there is quantum correlation. Interestingly, it appears there is similar behaviour in the expected payoff, however this is likely due to the choice of function, as it is not a universal feature. } \label{fig:1}
\end{figure*}

It is now beneficial to consider specific states which exhibit quantum discord, and study how the discord witness varies relative the expected payoff. Since both the expected payoff and discord witness depend on the measurements of both players, it is crucial to ensure the measurements found from the optimisation are the same when comparing the functions to ensure consistency. For this paper, all optimisation was performed numerically using Mathematica. To have a clear understanding of the relation between the two functions, both are maximised and minimised and compared against each other. The results for varying states are shown in Figs. 1 and 2.

Initially the Werner state \cite{Werner:1989aa} is considered, which is given by
\begin{equation}
\rho_W (\eta) = \frac{\mathbb{1}- \eta}{4} + \eta \ket{\psi_{-}} \bra{\psi_{-}},
\end{equation}
where $\ket{\psi_{-}} = (1/\sqrt{2})(\ket{\uparrow \downarrow} - \ket{\downarrow \uparrow})$, as this state is known to be classically correlated for $\eta=0$, discorded but non-entangled in the region $0\leq \eta \leq 1/3$, and entangled otherwise. 

A different discorded state which exhibits no entanglement is given by
\begin{equation}
\rho_D (x) = \frac{1}{2} \Big[ \ket{x} \bra{x} \otimes \ket{\uparrow} \bra{\uparrow} + \ket{\uparrow} \bra{\uparrow} \otimes \ket{x} \bra{x}\Big],
\end{equation}
where $\ket{x} = \cos x \ket{\uparrow} + \sin x \ket{\downarrow}$. This is clearly a mixed separable state, which has classical correlation at $x=0,\pi$, and has quantum discord otherwise. The relation between the expected payoff and the discord witness for these two states is shown in Fig. \ref{fig:1}.

\begin{figure*}[htp]
  \centering
  \subfloat[a][The expected payoff and discord witness for the Werner state, when the discord witness is maximised.] {\includegraphics[width=0.47\linewidth]{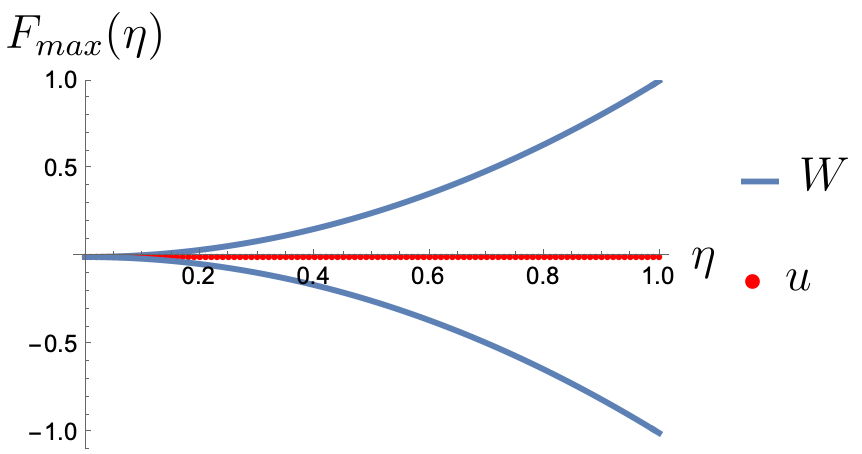}} 
  \label{fig:a} \subfloat[a][The expected payoff for the Werner state, when the discord witness is minimised.] {\includegraphics[width=0.47\linewidth]{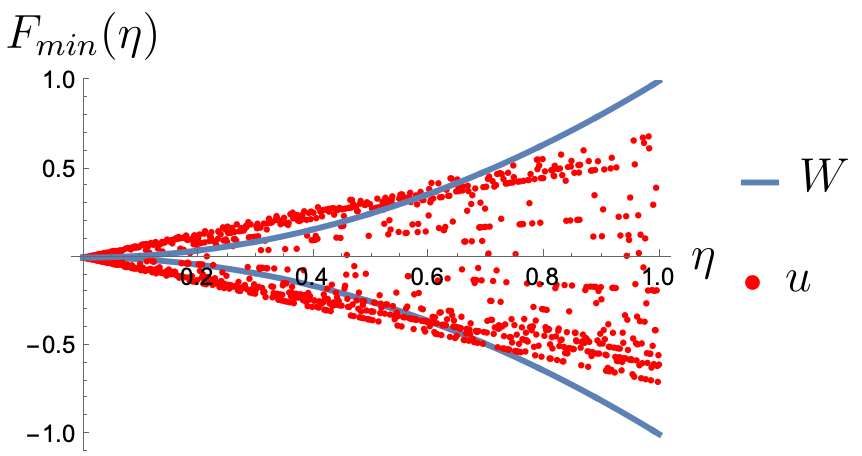}} 
  \label{fig:g} \\\subfloat[a][The expected payoff and discord witness for $\rho_D(x)$, when the discord witness is maximised.] {\includegraphics[width=0.47\linewidth]{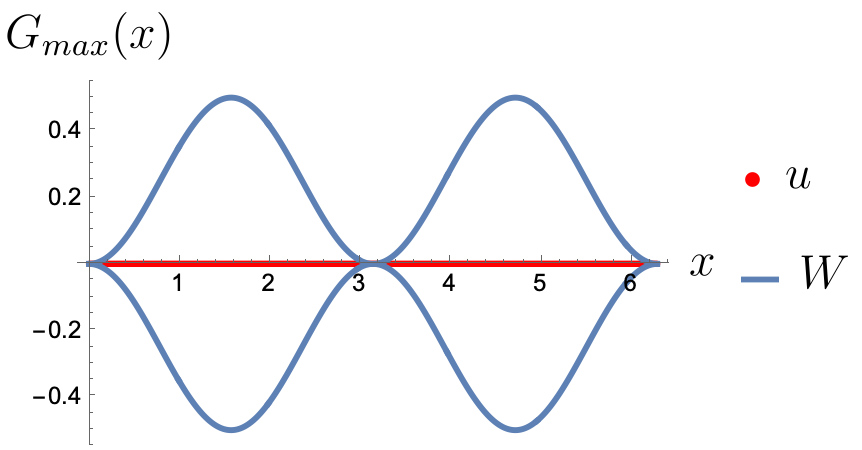}} 
  \label{fig:g} 
 \subfloat[a][The expected payoff and discord witness for $\rho_D(x)$, when the discord witness is minimised.]{\includegraphics[width=0.47\linewidth]{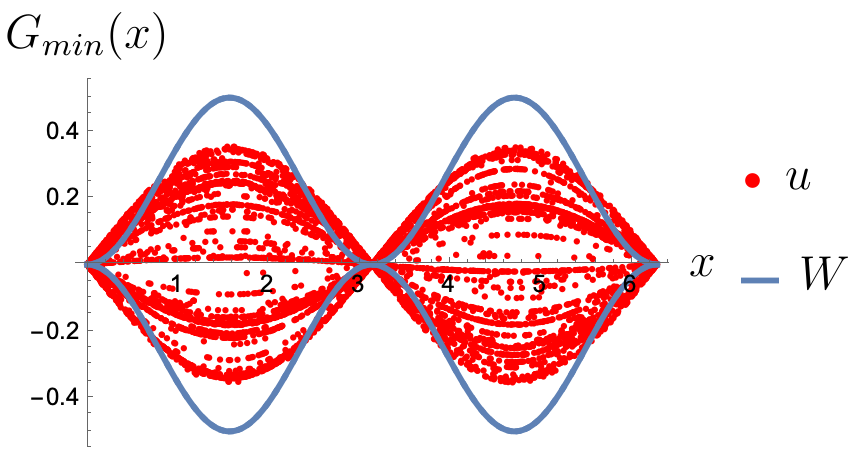}} \\
  \caption{This figure shows the results for the expected payoff and discord witness, when the discord witness has been both maximised and minimised. Interestingly, for these states the expected payoff either exhibits similar behaviour to the discord witness, or it is zero. To clarify, the results of the measurements from the optimisation is substituted into both functions. Additionally, to highlight the similarities between the discord witness and expected payoff, the negative of the discord witness is plotted alongside the maximised/minimised result.} \label{fig:2}
\end{figure*}

To clarify the notation, $F(\eta)$ denotes the set of the maximised expected payoff function and the discord witness, namely $F(\eta) = \{u(\eta), W(\eta)\} \equiv \{u, W\}$. Since the expected payoff has been maximised, the measurement values which maximise the expected payoff are then substituted into the discord witness. Conversely $G(x)$ is the same set and optimisation but now using $\rho_D(x)$ instead of the Werner state such that $G(x) = \{u(x), W(x)\} \equiv \{u, W\}$. It is crucial to emphasise each function is calculated for the same set of measurements which are found from the numerical maximisation of the expected payoff function. It is clear from Fig. \ref{fig:1}, that both the expected payoff and discord witness are zero when the quantum correlation is zero. This further demonstrates that the discord is a reliable witness of non-zero quantum discord. It is interesting to note that as the expected payoff increases, the discord witness decreases. However, given the discord witness is designed to detect quantum correlation due to it being non-zero, it is not concerned with the sign of the function. Therefore, it is conceivable to change the sign of the discord witness to ensure the witness is monotonic with the expected payoff. Whilst this is not considered in detail in this paper, it is an interesting question whether the sign of the discord witness has physical relevance, or whether it could be neglected and the absolute value of the discord witness could be used in the future.

In comparison, Fig. \ref{fig:2} considers the maximisation and minimisation of the discord witness, and the measurement values which satisfy this optimisation are substituted into the expected payoff. It is seen that when the discord witness is maximised, the expected payoff is zero. Interestingly, when the discord witness is minimised, the expected payoff exhibits similar qualitative behaviour, fluctuating between a positive and negative expected payoff. This is because it is a functional, so this behaviour is due to the distribution nature of result. To explicitly highlight this, the negative of the maximised/minimised discord witness is also included in the plots. Whilst the maximised/minimised discord witness exhibits a similar qualitative shape, it does not bound the distribution. It is an interesting question to understand the nature of this bound, and what physical relevance it has when determining quantum correlation.

\section{Discussion}
It is clear that there is an intrinsic link between Bayesian game theory and quantum discord as demonstrated by both analytical derivation and clear results from numerical examples by combining the CHSH game and the quantum discord witness. Despite this, there are still clear future directions of research which must be investigated to fully understand how quantum discord can be witnessed and quantified. For example, whether this link can be exploited and utilised to witness quantum advantage remains an open question, as there are currently few useful quantum protocols which incorporate quantum discord. Given the practical link between Bayesian game theory and useful implementable algorithms \cite{Khan:2018aa,Deligiannis:2017aa,Roy:2016aa}, it is expected this link could be exploited but must be rigorously demonstrated.

As discussed previously, finding an entirely general relation between Bayesian game theory and witnessing quantum discord will be of fundamental and practical relevance for the development of quantum technologies. Given there are assumptions about the nature of the game which are made when assigning payoffs in order to convey meaning to the game, it may be difficult to determine a generic relation. However, it should be investigated as a general relation would allow ease of mapping towards a variety of practically relevant protocols.

A unique feature of quantum discord relative to non-locality is the asymmetry that discord exhibits. This paper does not investigate whether an experimental witness could be used to determine and quantify this asymmetry. However, this should be investigated in the future as this would be a key departure from the standard CHSH game approach which does not explicitly study the asymmetry. Such a protocol could be used as a mechanism to determine one-way quantum advantage.

It is an interesting question to further determine whether a necessary and sufficient discord witness can be found and embedded into a Bayesian game-theoretic framework. This would enable a clear witness for determining quantum discord with the same robustness and clarity as the CHSH inequality. An example is shown in the appendix which demonstrates the limitations of this theory. Developing the theory further to be truly necessary and sufficient remains a key focus for future research despite the work presented here, as this would allow a universal test for quantum discord.

From the game-theoretic perspective, it would be an interesting question to determine whether the discord witness can be linked and utilised in a combative game, where the payoffs are no longer the same for each player. This has been studied when using entangled states \cite{Pappa:2015aa}, although this found the cooperative equilibria rather than the Nash equilibria. Subsequently, the relation between quantum discord and Nash equilibria is currently not understood in adversarial Bayesian games. Additionally, incorporating the prior beliefs further into the analysis may accurately reflect real-world scenarios allowing for future quantum protocols to be developed which can be utilised in real-world scenarios.

\section{Conclusion}
This paper has demonstrated a clear analytical link between Bayesian game theory and a quantum discord witness, where key examples have been considered and analysed. Subsequently, there is a clear operational framework which allows the detection of quantum discord, which can be mapped over to real-world game-theoretic scenarios, allowing for practical relevance and demonstration. This further highlights the relevance of quantum correlations in future quantum networks when looking to utilise and incorporate quantum effects.

\section{Acknowledgements}
AL is grateful to Rong Wang for insightful discussion on the discord witness, and how it can be utilised experimentally. 

\appendix*
\section{Zero-discord witness}
It is interesting to consider the scenario when the discord witness does not correctly determine quantum discord. Since the witness is sufficient, but not necessary there are certain states which do not exhibit a non-zero witness when there is quantum correlation. One such state is given by
\begin{equation}
\rho (x) = \frac{1}{2} \Big[ \ket{\uparrow} \bra{\uparrow} \otimes \ket{\uparrow} \bra{\uparrow} + \ket{x} \bra{x} \otimes \ket{x} \bra{x}\Big],
\end{equation}
\begin{figure}
\includegraphics[scale=0.6]{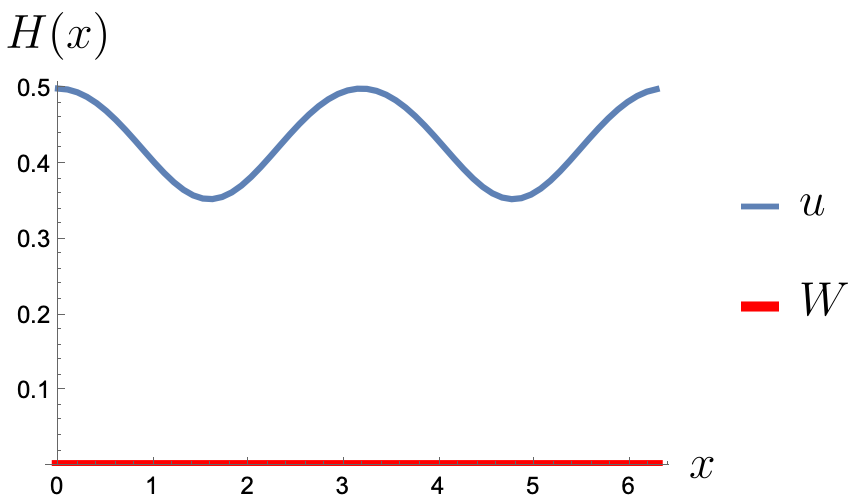}
\caption{This figure shows the relation between the expected payoff and discord witness when the discord witness fails to capture the quantum correlation. Subsequently, it is clear the previous links in Figs. 1 and 2 are not universal.}
\label{fig:3}
\end{figure}
where $\ket{x}$ is defined as before. For this state, it has non-zero quantum correlation everywhere except for $x=0,\pi$ on a $2\pi$ interval. It is clear from Fig. \ref{fig:3} that there is no clear relation between the discord witness and the maximised expected payoff. This is expected as since the discord witness is zero and therefore is independent of any choice of measurement, there will be no clear relation between the two functions. Moreover, the expected payoff appears to have maxima at zero-discorded values of $x$, demonstrating the limitations of this theory. This example demonstrates why understanding the general relation between quantum correlation, the expected payoff, and discord witness remains an interesting open question. In particular, understanding the work presented here for general states would also allow insight into the detection of quantum correlation, specifically in determining a necessary and sufficient condition.

\input{Draft.bbl}
%\bibliography{Literature_Review_References}% Produces the bibliography via BibTeX.

\end{document}